\documentclass[conference]{IEEEtran}
\IEEEoverridecommandlockouts

\usepackage{cite}
\usepackage{amsmath,amssymb,amsfonts}
\usepackage{algorithmic}
\usepackage{graphicx}
\usepackage{textcomp}
\usepackage[dvipsnames]{xcolor}
\usepackage{listings}
\usepackage[skins]{tcolorbox}
\usepackage{framed}
\usepackage{xurl}
\usepackage[hidelinks]{hyperref}
\usepackage{cleveref}
\usepackage{xurl}
\usepackage{tabularx}
\usepackage{colortbl}
\usepackage{booktabs}
\usepackage{multirow}
\usepackage{ascmac}
\usepackage{enumitem}
\usepackage{cprotect}
\usepackage{subcaption}
\usepackage{xspace}
\usepackage{tabularx}
\usepackage{booktabs}
\usepackage{mdframed}
\usepackage{fontawesome5}
\usepackage{ragged2e}
\usepackage{pifont}

\crefname{figure}{Figure}{Figure}
\crefname{table}{Table}{Table}
\crefname{section}{Section}{Sections}
\crefname{subsection}{Section}{Sections}

\newcommand{\TODO}[1]{{\textcolor{red}{#1}}}

\newcommand{\ReplicationURL}{\url{https://github.com/kusumotolab/quantum-defect-reproducibility}}

\newcommand{\symptom}{\faExclamationTriangle\xspace}
\newcommand{\patch}{\faTools\xspace}
\newcommand{\code}{\faCode\xspace}
\newcommand{\implication}[1]{\noindent \faHandPointRight[regular]\xspace \emph{#1}}

\newcommand{\ResearchUseBugsforQ}{li2024tosem,guo2024qse,tan2025fse,sato2025tse,ishimoto2025ease,yoshida2026saner}

\newcommand{\cmark}{\textcolor{ForestGreen}{\ding{51}}}  
\newcommand{\xmark}{\textcolor{red}{\ding{55}}}  

\newcommand{\bqp}{Bugs4Q-Robust\xspace}

\newtcolorbox{rqbox}{
    colback=gray!10,
    colframe=black,
    boxrule=0.4pt,
    sharp corners
}

{\endMakeFramed}

\definecolor{codebg}{RGB}{242,242,242}
\definecolor{codeline}{RGB}{170,170,170}
\definecolor{codecomment}{RGB}{0,128,0}
\definecolor{codekeyword}{RGB}{0,128,128}
\definecolor{codefunc}{RGB}{52, 148, 186}
\definecolor{codeop}{RGB}{0,128,128}

\lstdefinestyle{qiskitpython}{
    language=Python,
    basicstyle=\linespread{1.08}\ttfamily\footnotesize,
    backgroundcolor=\color{codebg},
    numbers=left,
    numberstyle=\tiny\color{codeline},
    numbersep=6pt,
    xleftmargin=1.2em,
    frame=leftline,
    rulecolor=\color{codeline},
    framesep=4pt,
    columns=fullflexible,
    keepspaces=true,
    showstringspaces=false,
    breaklines=true,
    tabsize=4,
    aboveskip=2pt,
    belowskip=2pt,
    keywordstyle=\color{codekeyword},
    commentstyle=\color{codecomment},
    stringstyle=\color{black},
    emph={execute,transpile,run},
    emphstyle=\color{codefunc},
    morekeywords={from,import},
}

\begin{document}

\title{On the Reproducibility of Quantum Software\\Defect Datasets: A Case Study of Bugs4Q}
\author{
    \IEEEauthorblockN{
        Haruto Ohto\IEEEauthorrefmark{1},
        Yuta Ishimoto\IEEEauthorrefmark{1},
        Shinsuke Matsumoto\IEEEauthorrefmark{1},
        Shinji Kusumoto\IEEEauthorrefmark{1}
    }
    \IEEEauthorblockA{
        \IEEEauthorrefmark{1}The University of Osaka, Osaka, Japan\\
        \{hr-ohto, ishimoto, shinsuke, kusumoto\}@ist.osaka-u.ac.jp
    }
}

\maketitle

\begin{abstract}
    The reproducibility of software defect datasets is essential for obtaining reliable and comparable research results.
    Zhu et al. have shown that defect datasets such as Defects4J suffer from reproduction failures (i.e., reported bugs become non-reproducible) as time passes since their creation.
    However, it remains unclear whether these findings generalize to \emph{quantum} software defect datasets.
    We therefore conduct a replication study of the prior work using Bugs4Q, a widely used dataset of real-world bugs in quantum programs.
    Our analysis includes 77{,}700 quantum program executions of 37 Bugs4Q artifacts across 21 core-library versions.
    The experimental results showed that the reproducibility of Bugs4Q dropped from 62.2\% on Qiskit v0.20.1 to 16.2\% on v2.3.1, the latest version as of April 1, 2026.
    A manual inspection of the root causes further indicated that 93.6\% of the failures were dependency-related.
    While these findings are consistent with those of the prior work, we also observed differences.
    In particular, most reproduction failures in Bugs4Q cannot be resolved merely by adjusting dependency versions; instead, they require source-code modifications such as migrating import paths and API invocations.
    Based on this observation, we curated \emph{\bqp}, a patched version of Bugs4Q to restore reproducibility.
    \bqp increases reproducibility from 16.2\% to 78.4\% on Qiskit v2.3.1.
    Our findings highlight the importance of continuous dataset maintenance in the rapidly evolving quantum software ecosystem.
\end{abstract}

\begin{IEEEkeywords}
    quantum software, software reproducibility, software defects, replication study
\end{IEEEkeywords}

\section{Introduction} \label{sec:intro}

Software defect datasets~\cite{just2014issta,le2015tse,tomassi2019icse,jimenez2023arxiv}, which consist of buggy programs and their corresponding fixed versions, have served as a fundamental resource for research on software testing~\cite{shamshiri2015ase,papadakis2018icse,perera2020ase}, fault localization~\cite{wong2016tse,kang2024fse,qin2025tse}, and automated program repair~\cite{monperrus2018csur,le2019cacm,huang2024csur}.
Reproducible defect datasets enable systematic evaluation and fair comparison of techniques.
This indicates that the reproducibility of defect datasets is essential for obtaining reliable and comparable research results.

Unfortunately, software defect datasets are not immune to \textit{reproduction failures} (i.e., reported bugs become non-reproducible) after their creation.
Zhu et al.~\cite{zhu2023icse} have shown that reproduction failures can occur even when the buggy and fixed code remains unchanged, because of changes in execution environments and system dependencies installed via package managers such as APT.
Even if a study is valid as long as the bugs were reproducible at the time of evaluation, subsequent reproduction failures hinder follow-up studies that seek to verify or extend its results.

While the prior work~\cite{zhu2023icse} has revealed reproduction failures in defect datasets for classical programs (i.e., programs intended to run on non-quantum computers), it remains unclear whether similar failures occur in \emph{quantum programs}. 
Unlike classical programs, quantum programs consist of quantum gates that manipulate qubits, the basic units of quantum computation~\cite{nielsen2010quantum}.
Driven by the growing need for quantum software engineering~\cite{zhao2020arxiv,leite2025tosem}, researchers have constructed several defect datasets for quantum programs~\cite{luo2022saner,zhao2023jss,usandizaga2026arxiv}.
Prior studies have reported quantum-specific categories of code-quality issues in quantum programs, including code smells~\cite{chen2023icse,stefano2024ese}, technical debt~\cite{openja2022jss,ishimoto2024apsec}, and bug patterns~\cite{zhao2021qse,luo2022saner,paltenghi2022oopsla}.
These quantum-specific characteristics suggest that findings on classical defect dataset reproducibility may not directly generalize to their quantum counterparts.

In this study, we conduct an \textit{operational replication}~\cite{gomez2014ist} of the study by Zhu et al.~\cite{zhu2023icse}.
That is, we retain the core objective of investigating defect dataset reproducibility while changing the target population to quantum defect datasets.
We focus on Bugs4Q~\cite{zhao2023jss}, a defect dataset widely used in research on testing and debugging quantum programs~\cite{\ResearchUseBugsforQ}.
Our study includes 77{,}700 quantum program executions of 37 Bugs4Q artifacts across 21 Qiskit core-library versions spanning three major Qiskit series released over a three-year period.
We use three reproducibility criteria adapted from the existing study~\cite{zhu2023icse}.
These criteria range from simply checking whether the expected buggy/fixed behavior is observed to stricter checks of whether the underlying cause of buggy behavior is consistent with the original source.
After this quantitative evaluation, we manually classify the root causes of reproduction failures.

Our study shows two findings consistent with those of the prior work.
First, the reproducibility of Bugs4Q degrades as time passes since its creation.
In our study, it drops sharply (i.e., from 62.2\% to 21.6\%) after major Qiskit version upgrades.
Second, dependency-related causes constitute the majority of reproduction failures (93.6\%), whereas quantum-specific causes account for only 5.1\%.
For example, many reproduction failures arise from mismatches between source-code references and updated Qiskit interfaces (e.g., artifacts fail with \texttt{ImportError} before execution reaches the original buggy statements).
We also observe a difference from classical defect datasets: resolving reproduction failures in Bugs4Q requires source-level patches (e.g., migrating stale Qiskit interface references), rather than environment-level patches such as pinning dependency versions.
Based on this observation, we curate \emph{\bqp}, a patched version of Bugs4Q designed to restore reproducibility.
\bqp increases reproducibility from 16.2\% to 78.4\% on Qiskit v2.3.1, the latest version as of April 1, 2026.
We release \bqp as a part of our replication package (see \cref{sec:data_availability}).

The contributions of this study are as follows:
\begin{itemize}
    \item We present the first empirical study of the reproducibility of a quantum software defect dataset over time.
    \item We classify the root causes of reproduction failures and show that dependency-related issues are dominant.
    \item We compare our findings with those on classical software defect datasets and discuss which findings do and do not generalize to quantum software defect datasets, deriving implications for maintaining such datasets.
    \item Leveraging our findings, we curate \bqp, a patched version of Bugs4Q that restores reproducibility on the latest version of Qiskit.
\end{itemize}

\textbf{Paper Organization.}
\cref{sec:setup} introduces our experimental design, including the research questions (RQs), dataset, reproducibility criteria, and experimental procedure.
\cref{sec:evaluation} presents the results and answers the RQs.
\cref{sec:bugs4q_plus} reports on the construction of \bqp\ and its reproducibility.
\cref{sec:discussion} discusses the extent to which we replicate the findings of Zhu et al.~\cite{zhu2023icse}\ and the implications drawn from our study.
\cref{sec:threat} discusses threats to validity, and \cref{sec:related_work} reviews related work.
\cref{sec:conclusion} concludes the paper.
\section{Experimental Design}
\label{sec:setup}

\begin{figure*}[t]
    \centering
    \includegraphics[width=.85\linewidth]{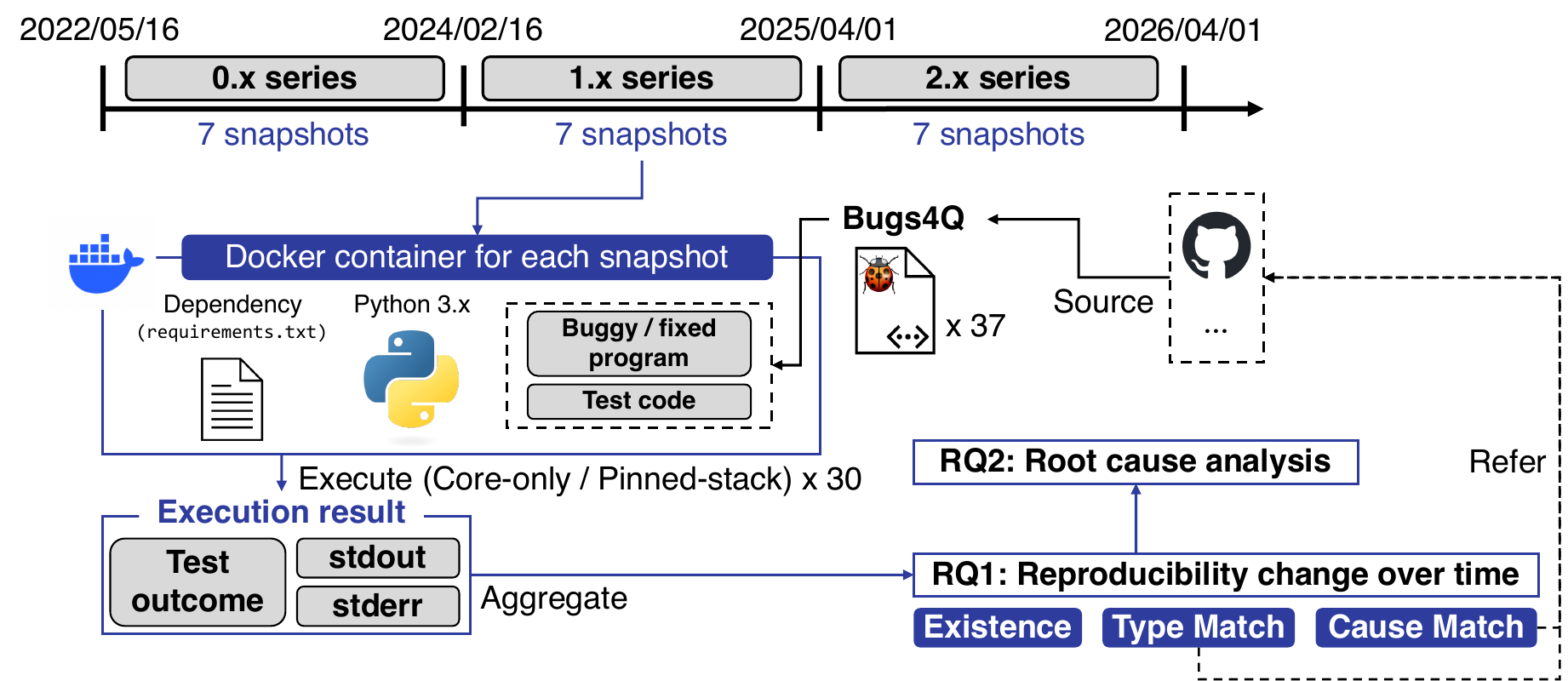}
    \caption{An overview of our experiment.}
    \label{fig:overview}
\end{figure*}

This section describes our experimental design, including the RQs, dataset, reproducibility criteria, and experimental procedure for our longitudinal analysis.

\subsection{Research Questions}
We address the following RQs, which are aligned with those of the prior work~\cite{zhu2023icse}:

\begin{description}
    \item[RQ1] \textit{How does the reproducibility of Bugs4Q change over time?}
    \item[RQ2] \textit{What are root causes for the reproduction failures of Bugs4Q?}
\end{description}

In RQ1, we quantitatively investigate the changes in reproducibility of Bugs4Q as the Qiskit ecosystem evolves,
because subsequent studies can rely on it as a benchmark only if its artifacts continue to exhibit the originally reported behaviors.
In RQ2, we qualitatively investigate why the reproduction failures occur.
While RQ1 reveals changes in reproducibility over time, it does not explain the underlying causes of reproduction failures; identifying these causes clarifies the maintenance required to keep Bugs4Q usable.

\subsection{Dataset}
\label{sec:dataset}

We use \textit{Bugs4Q}~\cite{zhao2023jss}, a real-world defect dataset for quantum programs written in Qiskit~\cite{javadi2024arxiv}.
Qiskit is a Python framework developed by IBM and the open-source community, with 7.3k stars on GitHub.\footnote{\url{https://github.com/Qiskit/qiskit}, as of May 7, 2026.}
Bugs4Q consists of 42 buggy Qiskit programs collected from three major platforms: GitHub, Stack Overflow, and Stack Exchange.
For each buggy program, Bugs4Q provides a corresponding fixed version and a test code that reproduces the bug.

To the best of our knowledge, Bugs4Q is the only benchmark that meets all three of the following criteria, which motivates our choice.
\begin{itemize}
    \item The buggy programs are written by developers; that is, the dataset contains real-world bugs.
    \item Both buggy and fixed programs are provided.
    \item Test code for reproducing the bugs is included.
\end{itemize}
Our criteria are aligned with those of Java defect datasets used in prior work~\cite{zhu2023icse}, such as Defects4J~\cite{just2014issta}.
The first criterion is necessary because we are interested in problems actually faced by programmers.
We therefore excluded synthetic datasets such as those based on quantum circuit mutations~\cite{usandizaga2026arxiv}.
The fixed programs are necessary to verify the expected behavior.
Test code is necessary to confirm that bugs in the buggy programs remain reproducible and that the expected behavior of the fixed programs is preserved over time.

We excluded five artifacts that lack test code, leaving 37 artifacts for our study.
Following the bug categories in the original paper~\cite{zhao2023jss}, these 37 artifacts are classified into one of the three categories.
\emph{Wrong Output (WO, 17 artifacts)} refers to artifacts where the output of the buggy program does not match the expected output specified by the test code.
\emph{Throw Exception (TE, 18 artifacts)} refers to artifacts where the buggy program raises an exception during execution.
\emph{Simulation Failure (SF, 2 artifacts)} refers to artifacts where the simulation fails for the buggy program.
Specifically, the simulation outcome is inspected by the \texttt{success} attribute of the object representing the simulation result.

\subsection{Criteria for Reproducibility}
\label{sec:criteria}

We assess the reproducibility of Bugs4Q artifacts using three criteria of increasing strictness: \emph{Existence}, \emph{Type Match}, and \emph{Cause Match}.
These criteria are adapted from the four reproducibility criteria used by the prior work~\cite{zhu2023icse}: \emph{Existence}, \emph{Number Match}, \emph{Name Match}, and \emph{Status Match}.
Since each Bugs4Q artifact has only a single test case, Number Match and Name Match are already captured by Existence.
Status Match, which checks the continuous integration (CI) build status, is not applicable to Bugs4Q because its artifacts are standalone scripts rather than projects with CI builds.
However, we preserve the underlying intent of Status Match, namely verifying that the overall execution state matches the record, by decomposing it into two levels of granularity: the type of the failure (i.e., Type Match) and its cause (i.e., Cause Match).

\textbf{Existence.}
It requires that the buggy version fails in any way and the fixed version succeeds.
Here, success is defined according to the bug type described in \cref{sec:dataset}, such as the program output for WO, normal termination for TE, or the simulation result for SF.
We count an artifact as non-reproducible whenever the fixed version does not succeed, even if the buggy version fails.
This treatment is consistent with the Existence criterion used in the prior work~\cite{zhu2023icse}.

\textbf{Type Match.}
It additionally requires that the observed failure type matches the one described in the original source of each Bugs4Q artifact, such as the corresponding Stack Overflow post.
For artifacts categorized as TE, Type Match is satisfied when the buggy version raises the same exception class as the one described in the source.
For artifacts categorized as WO, the expected behavior is encoded as assertions in the test code.
Therefore, Type Match is satisfied when the buggy version fails with an \texttt{AssertionError} raised by those assertions.
If the test fails due to an unrelated error, such as an \texttt{ImportError}, the artifact does not satisfy Type Match.
For artifacts categorized as SF, Type Match is satisfied when the simulation result indicates failure.

\textbf{Cause Match.}
It additionally requires that the runtime evidence is consistent with the cause described in the original source of each Bugs4Q artifact.
For artifacts categorized as TE, Cause Match is satisfied when the error message in the traceback is consistent with the cause described in the source.
For artifacts categorized as WO, Cause Match is satisfied when the incorrect output observed at runtime agrees with the incorrect output described in the source.
For example, if the source describes biased measurement counts toward a specific state, we regard the output as consistent when the reproduced buggy version exhibits the same bias.
For artifacts categorized as SF, Cause Match is satisfied when the simulator failure message\footnote{For this category, the message does not appear as a Python traceback because the buggy programs terminate normally.} is consistent with the cause described in the source.
Because semantic consistency in this criterion is difficult to check automatically, the first author manually validated it.

\subsection{Longitudinal Analysis}
\label{sec:procedure}

\cref{fig:overview} shows an overview of our experiment.
We study reproducibility across 21 snapshots from May 16, 2022, the release date of Bugs4Q, to April 1, 2026.
We use the release date as the starting point because we are interested in how reproducibility changes after the dataset became available to the research community.
Each snapshot simulates the environment at a given point in time by using the latest stable version of Qiskit available on PyPI by that date.
The snapshots cover three major Qiskit version series, namely 0.x, 1.x, and 2.x, with seven snapshots for each series.

\begin{table}[t]
    \centering
    \caption{Core-library versions studied in this paper.}
    \label{tab:versions}
    \setlength{\tabcolsep}{4pt}
    \renewcommand{\arraystretch}{1.1}
    \begin{tabularx}{\columnwidth}{
        >{\centering\arraybackslash}p{0.10\columnwidth}
        >{\RaggedRight\arraybackslash}p{0.50\columnwidth}
        >{\RaggedRight\arraybackslash}p{0.32\columnwidth}
        }
        \toprule
        Series                 & Versions & Period \\
        \midrule
        0.x                    &
        0.20.1 /  0.21.1 /  0.21.2 /  0.22.2 / 
        0.22.3 /  0.23.2 /  0.24.0 &
        2022/05/16--2023/05/16                     \\
        1.x                    &
        1.0.0 /  1.0.2 /  1.1.0 /  1.2.0 / 
        1.2.4 /  1.3.1 /  1.3.2    &
        2024/02/16--2025/02/16                     \\
        2.x                    &
        2.0.0 /  2.0.2 /  2.1.1 /  2.2.1 / 
        2.2.3 /  2.3.0 /  2.3.1    &
        2025/04/01--2026/04/01                     \\
        \bottomrule
    \end{tabularx}
\end{table}

\textbf{Qiskit Versions and Snapshot Periods.}
\cref{tab:versions} lists the Qiskit versions and snapshot periods for each series.
For the 0.x series, the first snapshot is taken on the release date of Bugs4Q.
For the 1.x and 2.x series, the first snapshot is taken on the release date of the initial version of each series, because we aim to evaluate reproducibility after each major version upgrade.
In all series, subsequent snapshots are taken every two months over the following year, yielding seven snapshots per series.
We use Python 3.9.25 for the 0.x and 1.x series, and Python 3.10.20 for the 2.x series.
In total, these 21 snapshots cover a three-year observation period.

\begin{figure*}[t]
    \centering

    \begin{subfigure}{\linewidth}
        \centering
        \includegraphics[width=0.9\linewidth]{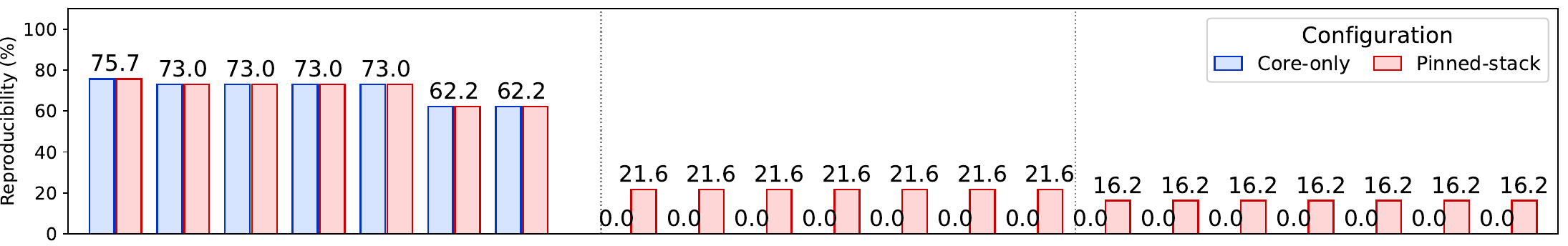}
        \caption{Existence}
        \label{fig:existence}
    \end{subfigure}

    \begin{subfigure}{\linewidth}
        \centering
        \includegraphics[width=0.9\linewidth]{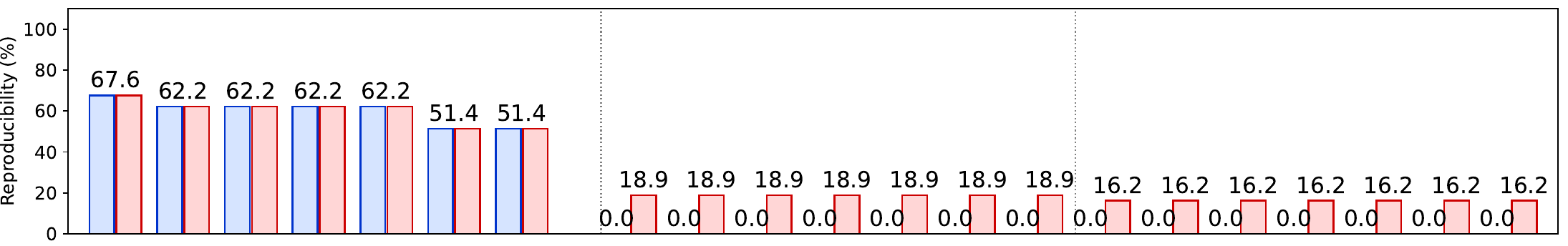}
        \caption{Type Match}
        \label{fig:type}
    \end{subfigure}

    \begin{subfigure}{\linewidth}
        \centering
        \includegraphics[width=0.9\linewidth]{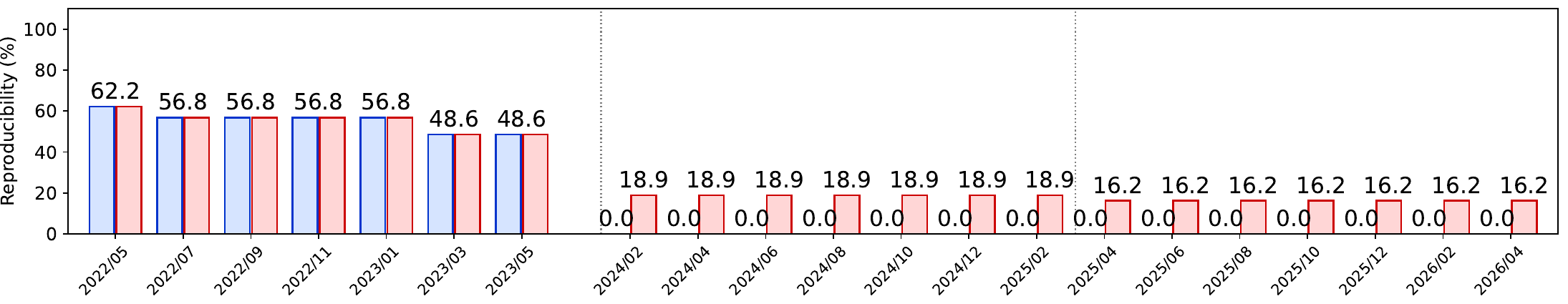}
        \caption{Cause Match}
        \label{fig:cause}
    \end{subfigure}

    \caption{
        Reproducibility for each snapshot under the three criteria.
        The value above each bar indicates the percentage of reproducible artifacts at the corresponding snapshot; when a single value is shown, both configurations have the same percentage.
        Dashed vertical lines separate the major versions of Qiskit: 0.x, 1.x, and 2.x series.
    }
    \label{fig:reproducibility_by_criterion}
\end{figure*}

\textbf{Environment at Each Snapshot.}
For each snapshot, we build a separate Docker image to isolate the corresponding execution environment.
For each Docker container, we evaluate each Bugs4Q artifact under two dependency configurations: \emph{Core-only} and \emph{Pinned-stack}.
The \emph{Core-only} configuration represents a scenario in which a user updates only the core Qiskit library (i.e., \texttt{qiskit-terra} for the 0.x series and \texttt{qiskit} for the 1.x and 2.x series) to the target version without accounting for the surrounding ecosystem.
In contrast, the \emph{Pinned-stack} configuration represents a scenario in which a user familiar with the Qiskit ecosystem consults the official Qiskit release notes and documentation when selecting package versions compatible with the core library.
Specifically, this configuration updates the core Qiskit library and, guided by the official documentation, pins compatible versions of surrounding packages and removes incompatible ones.
This configuration captures a practical scenario because, for example, the official Qiskit 1.0 documentation\footnote{\url{https://quantum.cloud.ibm.com/docs/en/guides/qiskit-1.0-installation}} states that ``\textit{Even more unfortunately, we cannot communicate this conflict to pip because of limitations in its metadata system,}'' suggesting that incompatibilities in the Qiskit package ecosystem may require manual intervention.

\textbf{Execution at Each Snapshot.}
We execute the buggy and fixed versions of each artifact through the bundled test case on the corresponding Docker container.
Because quantum programs may produce non-deterministic outputs~\cite{zhang2023esem,kaur2025saner}, we run both versions 30 times, following empirical guidelines~\cite{arcuri2011icse}.
For each execution, we record the test outcome and save the runtime log, including stdout and stderr.
We then check whether each of the three reproducibility criteria defined in \cref{sec:criteria} is satisfied.

\section{Evaluation}
\label{sec:evaluation}

All our experiments are performed on a classical computer (i.e., a non-quantum computer) using the simulator functionality provided by Qiskit.
Although programs written in Qiskit can be executed on actual quantum computers, we use the simulator due to the limited availability of real quantum hardware.

\subsection{RQ1: Bugs4Q Reproducibility Change}
\label{sec:rq1}

\textbf{Targets.}
At each snapshot defined in \cref{sec:procedure}, we execute the buggy and fixed quantum programs of each Bugs4Q artifact.
Our analysis covers 77{,}700 quantum program executions in total.
While the full combination of 21 snapshots, 2 dependency configurations, 37 artifacts, 30 repeated runs, and buggy/fixed programs would yield 93{,}240 executions, 7 out of 42 ($21 \times 2$ snapshot-configuration pairs) Docker image builds failed,\footnote{These build failures occur during the execution of \texttt{pip install -r requirements.txt} in the Dockerfile, i.e., when the Python version and the dependencies specified in \texttt{requirements.txt} cannot coexist.} reducing the actual number of executions.

\textbf{Evaluation Metrics.}
For each reproducibility criterion $C \in \{\text{Existence}, \text{Type Match}, \text{Cause Match}\}$, artifact $p$ is considered reproducible at snapshot $t$ if criterion $C$ is satisfied in all 30 trials.
This definition treats flaky artifacts~\cite{parry2021tosem,zhang2023esem,kaur2025saner} (i.e., those that intermittently succeed and fail) as non-reproducible.
We define $\text{Repro}_C(t,p)$ as an indicator variable that takes 1 when this condition holds and 0 otherwise.
The number of reproducible artifacts at snapshot $t$ under criterion $C$ is then:

\begin{equation}
    R_C(t) = \sum_{p} \text{Repro}_C(t, p).
\end{equation}
We report $R_C(t)$ as the percentage of reproducible artifacts for each snapshot $t$ and each criterion $C$ to examine how reproducibility changes over time.

\textbf{Results.}
\cref{fig:reproducibility_by_criterion} plots $R_C(t)$ for each criterion $C$ and each snapshot $t$, under both the Core-only and Pinned-stack configurations.

\textbf{As shown in \cref{fig:reproducibility_by_criterion}, Bugs4Q reproducibility dropped sharply across major Qiskit versions: from 48.6\%--75.7\% in the 0.x series to 0.0\%--21.6\% in the 1.x and 2.x series, indicating that it retains a meaningful level of reproducibility only under legacy versions.}
This sharp drop is consistent with the breaking changes introduced at each major Qiskit release~\cite{cardinal2025arxiv,suarez2025arxiv,qiskit2024migration,qiskit2025migration}.
Notably, even at the earliest snapshot, v0.20.1, 75.7\% of artifacts satisfied Existence.
This is lower than Defects4J, the dataset with the highest Existence reproducibility (96.9\%) reported by Zhu et al.~\cite{zhu2023icse}.
This imperfection is plausibly due to the non-determinism of quantum measurements as well as inconsistencies between the environment used during Bugs4Q construction and our experimental setup, since Bugs4Q does not record the exact library versions used at collection time.
We further investigate the root causes of these reproduction failures in RQ2.

\textbf{Pinned-stack mitigates the reproducibility collapse caused by Qiskit's major version upgrades, e.g., restoring Existence from 0.0\% to 21.6\% in the 1.x series.}
Across the 0.x series, Core-only and Pinned-stack reproduce the same set of artifacts at every snapshot.
In contrast, as shown in \cref{fig:reproducibility_by_criterion}, no Bugs4Q artifact is reproducible (0.0\%) from v1.0.0 onward under the Core-only configuration.
By adjusting the surrounding Qiskit-related packages, Pinned-stack restores Existence to 21.6\% in the 1.x series and 16.2\% in the 2.x series, and Cause Match to 18.9\% and 16.2\%, respectively.
This partial recovery is achieved by excluding legacy ecosystem packages (e.g., \texttt{qiskit-aqua} and \texttt{qiskit-ignis}) that cannot coexist with the v1.0.0 or later core library.

\textbf{Over the three-year observation period, 31 of the 37 artifacts (83.8\%) experienced reproduction failures at some snapshot even with Pinned-stack, indicating that they affect the majority of Bugs4Q.}
On the modern core-library versions in particular, Bugs4Q remains poorly reproducible (up to 21.6\% for the 1.x series and 16.2\% for the 2.x series), although Pinned-stack consistently yields better reproducibility than Core-only.
We thus use Pinned-stack as the basis for the following analyses.

\begin{figure}[t]
    \centering
    \includegraphics[width=\linewidth]{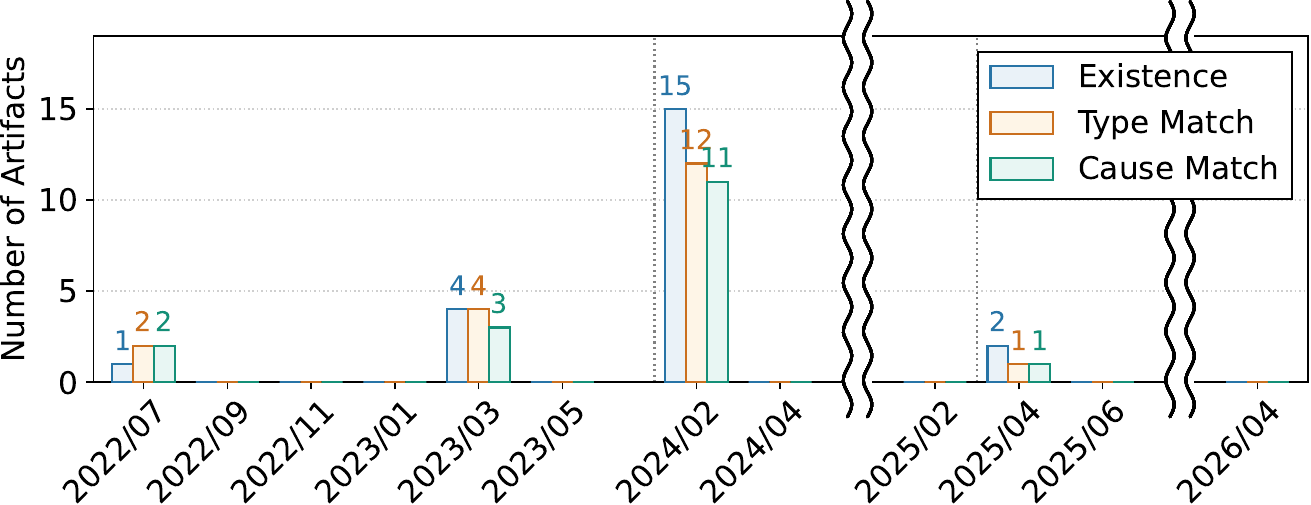}
    \caption{
    Number of newly broken artifacts for each criterion under Pinned-stack. 
    Dashed vertical lines separate the major versions of Qiskit: 0.x, 1.x, and 2.x series.
    Periods with no newly broken artifacts are abbreviated by wavy lines.
    }
    \label{fig:broken}
\end{figure}

\textbf{Once a Bugs4Q artifact becomes non-reproducible at a snapshot, it never returns to a reproducible state in any subsequent snapshot, suggesting that periodic maintenance is necessary to keep a defect dataset usable.}
Prior work on classical defect datasets~\cite{zhu2023icse} observed both newly broken and newly reproducible artifacts.
Here, an artifact is newly broken (newly reproducible) for a criterion if it does not satisfy (satisfies) that criterion at the current snapshot but satisfied (did not satisfy) it at the previous snapshot.
In our study, we observed newly broken artifacts but no newly reproducible artifacts.
\cref{fig:broken} shows the number of newly broken artifacts at each snapshot.
Newly broken artifacts tend to emerge at major version transitions (i.e., right after the dashed vertical lines in \cref{fig:broken}) rather than at minor version updates.

\begin{rqbox}
    \noindent
    \textbf{RQ1:}
    Bugs4Q reproducibility dropped sharply (e.g., from 62.2\% to 21.6\%) once major version upgrades of Qiskit occur after its construction.
    Updating both core and surrounding Qiskit-related packages (Pinned-stack) partially mitigates this collapse (e.g., +21.6 percentage points in Existence) but cannot restore reproducibility to the level observed in the 0.x series.
    Over the three-year observation period, 31 of the 37 Bugs4Q artifacts (83.8\%) experienced reproduction failures at least once.
\end{rqbox}

\subsection{RQ2: Root Causes and Patch Strategies}
\label{sec:rq2}

\textbf{Targets.}
We focus on artifacts that do not satisfy the strictest criterion, Cause Match, under the Pinned-stack configuration.
This selection covers all reproduction failures, since artifacts that violate the two weaker criteria also violate Cause Match.
Our analysis targets the set of snapshot--artifact pairs $(t,p)$ for which $\text{Repro}_{\text{Cause Match}}(t,p)=0$, totaling 543 pairs.

\textbf{Classification of Reproduction Failures.}
The reproduction failure categories proposed by Zhu et al.~\cite{zhu2023icse} are too specific to fit our setting, as they depend on particular Java build tools such as Maven and Gradle.
Therefore, we adopt an open coding approach~\cite{charmaz2006grounded} to classify reproduction failures in quantum software defect datasets, since no existing taxonomy is available for this purpose.
For each target (snapshot, artifact) pair, two of the authors independently classify the root cause of each of the 543 instances by inspecting the runtime logs and the Bugs4Q source (e.g., the linked GitHub issues from which the artifact was collected), assigning a short descriptive code.
After the independent coding is completed, we compute Fleiss's kappa~\cite{fleiss1971apa} to assess the inter-rater reliability.
The resulting value is 0.67, indicating \textit{substantial agreement}~\cite{viera2005fammed}.
The two annotators then jointly resolve disagreements through discussion to determine the final root cause of each instance, and refine the categories by merging semantically equivalent codes.
In addition, for each resulting root-cause category, we describe a patch strategy to restore reproducibility, which the annotators came up with during the coding process while inspecting the raw data such as the runtime logs.

\begin{table*}[t]
    \centering
    \caption{Root causes and possible patch strategies. Gray-shaded rows indicate the main categories. N/A indicates that no general patch strategy is applicable. Repro. denotes reproduction.}
    \label{tab:rq2-cause}
    \renewcommand{\arraystretch}{1.1}
    \begin{tabular}{p{0.217\linewidth} r p{0.318\linewidth} p{0.3\linewidth}}
        \toprule
        Root cause                                & Percentage                                                                                                                                                    & Definition of main category & Patch strategy \\
        \midrule

        \cellcolor{gray!20}
        \textbf{Source-related repro. failures}   & \cellcolor{gray!20}\textbf{82.0\%}
                                                  & \multirow[t]{3}{=}{\raggedright Failures caused by artifact source code using Qiskit interfaces that are no longer valid in the target Qiskit version.}
                                                  & \multirow[t]{3}{=}{\raggedright Update obsolete import paths or replace incompatible API invocations with calls supported by the target Qiskit version.}                                                     \\
        Import incompatibility                    & 48.1\%                                                                                                                                                        &                             &                \\
        API invocation incompatibility            & 33.9\%                                                                                                                                                        &                             &                \\
        \midrule

        \cellcolor{gray!20}
        \textbf{Library-related repro. failures}  & \cellcolor{gray!20}\textbf{11.6\%}
                                                  & \multirow[t]{4}{=}{\raggedright Failures caused by changes in Qiskit-related libraries, even though artifact source code still uses valid Qiskit interfaces.}
                                                  & \multirow[t]{4}{=}{\raggedright Adjust dependency versions or update test oracles when applicable; otherwise N/A.}                                                                                           \\
        Library--library incompatibility          & 4.6\%                                                                                                                                                         &                             &                \\
        Library-internal behavior change          & 6.1\%                                                                                                                                                         &                             &                \\
        Output format change                      & 0.9\%                                                                                                                                                         &                             &                \\
        \midrule

        \cellcolor{gray!20}
        \textbf{Quantum-specific repro. failures} & \cellcolor{gray!20}\textbf{5.1\%}
                                                  & Failures caused by probabilistic variation in quantum measurement or simulation results.
                                                  & Tune the shot count and relax the oracle threshold.                                                                                                                                                          \\
        \midrule

        \cellcolor{gray!20}
        \textbf{Original test issues}             & \cellcolor{gray!20}\textbf{1.3\%}
                                                  & Failures where the original test does not adequately validate program behavior, making the correctness of the fixed version unclear.
                                                  & N/A                                                                                                                                                                                                          \\

        \bottomrule
    \end{tabular}
\end{table*}

\textbf{Results.}
\cref{tab:rq2-cause} summarizes the root-cause categories, the corresponding patch strategies, and the percentage of instances in each category.
Our taxonomy consists of four main categories and five subcategories, yielding seven leaf categories in total.
The four main categories are \textit{Source-related reproduction failures} (82.0\%, caused by mismatches between artifact source code and updated Qiskit interfaces), \textit{Library-related reproduction failures} (11.6\%, caused by changes in Qiskit-related libraries),
\textit{Quantum-specific reproduction failures} (5.1\%, caused by quantum-specific probabilistic variation), and \textit{Original test issues} (1.3\%, caused by insufficient validation in the original test).
The rest of this section discusses each category in turn, presenting its symptoms (see \symptom icon), code examples when available (\code), and potential patch strategies (\patch).

\paragraph{Import incompatibility}
This is the most common root cause, accounting for 261 failures (48.1\%).
\symptom These failures are caused by import-path mismatches: the artifact source code remains unchanged, but the referenced modules or packages are no longer available at their original paths in the target Qiskit version.
\code The following code excerpt from an artifact (ID:19 in Bugs4Q) provides an example of this root cause.
\begin{lstlisting}[style=qiskitpython,numbers=none]
from qiskit.extensions import HGate, CXGate
\end{lstlisting}
This import statement fails in v1.0.0 and later because the legacy \texttt{qiskit.extensions} module is no longer available at that import path.
\patch The patch strategy updates such outdated import paths to their replacements documented in the Qiskit migration guides\cite{qiskit2024migration,qiskit2025migration}.
In the code excerpt above, the legacy \texttt{qiskit.extensions} import is migrated to \texttt{qiskit.circuit.library}.

\paragraph{API invocation incompatibility}
This root cause accounts for 184 failures (33.9\%).
\symptom These failures occur when the artifact source code invokes Qiskit APIs that have been removed or whose call forms have changed in the target Qiskit version.
\code The following code excerpt from an artifact (ID:39 in Bugs4Q) provides an example of this root cause.
\begin{lstlisting}[style=qiskitpython,numbers=none]
job = execute(
    qc,
    backend = Aer.get_backend('qasm_simulator'),
    shots=1024
)
\end{lstlisting}
This code submits a circuit to a backend through the legacy \texttt{execute} API function. The failure surfaces only when \texttt{execute} is called.
\patch The patch strategy replaces legacy API invocations with equivalents supported by the target version.

\paragraph{Library--library incompatibility}
This root cause accounts for 25 failures (4.6\%).
\symptom These failures occur when Qiskit-related packages lose runtime compatibility with each other, even though they can still be installed together.
\patch A patch strategy is to adjust the versions of the surrounding Qiskit-related packages so that they remain mutually compatible with the target Qiskit version.
This is the only root cause whose patch strategy operates entirely at the environment level (e.g., version pinning), without modifying artifact source code or test oracles.

\paragraph{Library-internal behavior change}
This root cause accounts for 33 failures (6.1\%).
\symptom These failures occur when the Qiskit interfaces used by the artifact remain valid, but changes in the internal behavior cause the reproduction failures.
For instance, an operation that originally triggered the bug may now be handled internally by Qiskit, causing the original bug-triggering condition to disappear.
In such cases, no general patch strategy is applicable.

\paragraph{Output format change}
This root cause accounts for five failures (0.9\%).
\symptom These failures occur when Qiskit changes the textual representation of outputs while the program behavior remains unchanged.
For example, a qubit label printed in the output changes from \texttt{q\_0} to \texttt{q}, causing a string-based assertion to fail.
\patch A patch strategy is to update the test oracle to match the output format of the target Qiskit version.
This modification is confined to the test oracle and does not alter the program logic.

\paragraph{Quantum-specific reproduction failures}
This root cause accounts for 28 failures (5.1\%).
\symptom These failures stem not from API or dependency changes, but from the probabilistic nature of quantum programs: the observed output may occasionally deviate from the expected distribution due to non-determinism of quantum programs, even when the source code is unchanged.
\code The following excerpt from an artifact (ID:26 in Bugs4Q) provides an example of this root cause.
\begin{lstlisting}[style=qiskitpython,language=Python,numbers=none,emph={}]
circuit = QuantumCircuit(2)
circuit.h(0)
circuit.x(1)
circuit.cx(0, 1)
circuit.measure_all()
backend = QasmSimulator()
job = backend.run(..., shots=1024)
counts = job.result().get_counts(circuit)
\end{lstlisting}
The excerpt executes a quantum circuit and returns sampled counts for each state from measurement.
The ideal output distribution assigns a probability of 0.5 to each of the \texttt{01} and \texttt{10} states.
In our setting of 1{,}024 shots, the expected count for each state is 512.
Thus, the corresponding test applies a statistical assertion~\cite{huang2019ics,sato2025tse} to check whether the sampled counts deviate from the ideal distribution.
\begin{lstlisting}[style=qiskitpython,language=Python,numbers=none]
# Test side
pvalue = compute_pvalue(counts, expected)
self.assertGreater(pvalue, alpha)
# STDERR
AssertionError: 0.0244... not greater than 0.05
\end{lstlisting}
In the observed run, the sampled counts were \texttt{\{'01': 476, '10': 548\}}.
Although the result appeared close to the expected distribution, the test oracle rejected it and caused the fixed version to fail.
\patch A patch strategy is to tune the shot count and oracle threshold so that statistically plausible sampling variation is not rejected as a reproduction failure, while preserving the oracle's ability to distinguish buggy and fixed behavior.

\paragraph{Original test issues}
This root cause accounts for the remaining seven failures (1.3\%), all originating from a single artifact.
\symptom These failures stem from a problem in the original test rather than Qiskit evolution: the test does not seem to correctly check the target behavior.
We could not confirm whether the fixed program produced the expected output when Bugs4Q was constructed.
Given this uncertainty about the correctness of the fixed program, we consider this case not patchable in this study.

We further observe that some artifacts contribute multiple reproduction failures classified under different root causes across Qiskit versions.
For example, an artifact (ID:20 in Bugs4Q) fails under \textit{Library-internal behavior change} in v0.20.1 -- v0.22.3, shifts to \textit{API invocation incompatibility} in v0.23.2 -- v0.24.0, and finally fails under \textit{Import incompatibility} from v1.0.0 and later.
This indicates that addressing one root cause for a given artifact does not guarantee its long-term reproducibility, because the same artifact may later experience failures caused by different root causes.

\begin{rqbox}
    \noindent
    \textbf{RQ2:}
    We identified seven root causes across 543 reproduction failures.
    Among these failures, 93.6\% are dependency-related (Source-related: 82.0\%, Library-related: 11.6\%).
    The most frequent leaf category is \textit{Import incompatibility} (48.1\%), where import statements reference Qiskit modules that have become incompatible.
    Only 4.6\% of failures have a patch strategy that does not require modifying artifact source code.
\end{rqbox}
\section{Improving the Reproducibility of Bugs4Q}
\label{sec:bugs4q_plus}

As an application of our findings from the previous RQs, we improve the reproducibility of Bugs4Q.

\subsection{Motivation}
Our investigation in the previous RQs revealed two key findings: (1) Bugs4Q is largely unreproducible under the latest Qiskit version (e.g., v2.3.1), and (2) some categories of root causes might be addressed by patch strategies.
Although Bugs4Q has been widely used in research on testing and debugging quantum programs~\cite{\ResearchUseBugsforQ}, some studies use only a small fraction of its 42 artifacts, such as 5 in~\cite{li2024tosem}, 8 in~\cite{sato2025tse}, and 18 in~\cite{yoshida2026saner}.
Thus, applying patches to restore its reproducibility under the latest Qiskit environment is essential to enable future research building upon this benchmark.
We refer to the patched dataset as \emph{\bqp}, which targets reproducibility under v2.3.1.

\subsection{Construction}
\label{sec:bugs4q_plus_construction}
We construct Bugs4Q-Robust by manually creating concrete patches guided by the patch strategies identified in RQ2.
Patches are applied only to root causes that have an applicable patch strategy; \emph{Library-internal behavior change} and \emph{Original test issues} are left unpatched.
For each artifact, we restrict modifications to the observed root cause and keep the patch minimal, preserving the original program logic and test intent.

\textbf{Environment-level Patches.}
We first apply patches that do not require source-code editing.
These patches address \emph{Library--library incompatibility} by adjusting the dependency specification.
Specifically, we select the latest versions of the surrounding Qiskit-related packages that remain mutually compatible with v2.3.1.
This is the only environment-level patch applied in \bqp.

\textbf{Code-level Patches.}
We next apply patches that modify artifact source code.
For \emph{Import incompatibility} and \emph{API invocation incompatibility}, we migrate the affected import statements and API invocations following the Qiskit migration guide~\cite{qiskit2024migration,qiskit2025migration}.
One common patch is the migration from the legacy \texttt{execute} API function to the recommended \texttt{transpile}--\texttt{backend.run} procedure as shown below.
\begin{lstlisting}[style=qiskitpython,numbers=none]
backend = Aer.get_backend('qasm_simulator')
new_circuit = transpile(qc, backend)
job = backend.run(new_circuit, shots=1024)
\end{lstlisting}
We applied this migration to 12 of the 37 artifacts.

\textbf{Data Availability.}
We release \bqp as a public benchmark under v2.3.1, included in our replication package (see \cref{sec:data_availability}).
The release contains all 37 artifacts used in this study.
Each artifact contains the buggy program, fixed program, test code, and metadata such as the original issue links from which the artifact was constructed.
At the repository root, we provide runtime logs, a runner script, dependency specifications, and auxiliary scripts for reconstructing and executing the benchmark.
This allows future users to rerun the benchmark and inspect both recovered and unrecovered artifacts under the same setup.

\subsection{Reproducibility}
\label{sec:bugs4q_plus_reproducibility}

\cref{tab:repro-bqp} shows the reproducibility of \bqp under v2.3.1, the latest version as of April 1, 2026.
For reference, this table also includes the reproducibility results of the original Bugs4Q under the v2.3.1 snapshot with the Pinned-stack dependency configuration.

\textbf{On average, the reproducibility of \bqp improved by 51.4 percentage points over the original Bugs4Q across the three criteria.}
This demonstrates the effectiveness of the patch strategies identified in RQ2.
The largest gain (+62.2 percentage points) was observed under Existence, indicating that the applied patches enable only the buggy version of many artifacts to reach a buggy line.
The gain was smaller under the stricter criteria (+48.7 percentage points for Type Match and +43.3 percentage points for Cause Match), reflecting that some restored failures do not exactly match the originally reported ones in terms of error type or root cause.

To investigate the causes of artifacts that remain non-reproducible in \bqp, the first author labeled them by reusing the categories established in RQ2.
\cref{tab:remain-cause} summarizes the classification results of the 15 artifacts that still fail to satisfy Cause Match.
\textbf{\cref{tab:remain-cause} shows that most failures categorized as \emph{Source-related reproduction failures} have been resolved in \bqp.}
The remaining four cases of \textit{Import incompatibility} involve Qiskit-related packages or modules that have been removed without direct replacements.
Overall, the dominant remaining cause is \emph{Library-internal behavior change}, which accounts for 10 of the 15 remaining artifacts.
This indicates that \bqp effectively resolves reproduction failures caused by source-level interface mismatches, whereas library-internal behavior changes remain difficult to recover through our applied patches.

\textbf{\textit{Library-internal behavior change} can eliminate the original bug-triggering conditions.}
For example, one artifact (ID:14 in Bugs4Q), which could not be reproduced even in \bqp, reported a bug caused by omitting the diagonalization step required when using \texttt{PauliExpectation}.
In the successor \texttt{Estimator} API, however, diagonalization is handled implicitly, so the same user-level omission no longer triggers the original bug.
This represents a fundamental limitation of bug reproduction: when the new API automatically handles what users previously had to do manually, the original bug can no longer be triggered, regardless of patching.

\begin{table}[tb]
    \centering
    \caption{Reproducibility of \bqp.}
    \label{tab:repro-bqp}
    \begin{tabular}{lccc}
        \toprule
        Dataset           & Existence       & Type Match      & Cause Match     \\
        \midrule
        \textbf{\bqp}     & \textbf{78.4\%} & \textbf{64.9\%} & \textbf{59.5\%} \\
        Bugs4Q (original) & 16.2\%          & 16.2\%          & 16.2\%          \\
        \bottomrule
    \end{tabular}
\end{table}
\begin{table}[t]
    \centering
    \caption{Remaining root causes of reproduction failures in \bqp.}
    \label{tab:remain-cause}
    \begin{tabular}{lrr}
        \toprule
        Root cause                       & \# Artifacts & Percentage \\
        \midrule
        Import incompatibility           & 4            & 26.7\%     \\
        Library-internal behavior change & 10           & 66.7\%     \\
        Original test issues             & 1            & 6.7\%      \\
        \bottomrule
    \end{tabular}
\end{table}
\section{Discussion}
\label{sec:discussion}

\begin{table*}[t]
  \centering
  \caption{Comparison of previous findings (\textbf{PF}) reported by Zhu et al.~\cite{zhu2023icse} on classical software defect datasets and our findings (\textbf{OF}) on the quantum software defect dataset Bugs4Q, together with their implications. The rightmost column indicates whether the previous and our findings are similar (\cmark) or different (\xmark).}
  \label{tab:replication-discussion}
  \renewcommand{\arraystretch}{1.25}
  \begin{tabular}{@{}p{0.075\linewidth} p{0.37\linewidth} p{0.43\linewidth} c@{}}
    \toprule
    Our RQ & Finding comparison                                                                                                                                                                                                                        & Implication & Match \\
    \midrule

    \multirow{2}{=}{\raggedright RQ1: Reproducibility over time}
           & \textbf{PF1:} \underline{62.6\%} of artifacts experience reproduction failures at least once. \newline
    \textbf{OF1:} \underline{83.8\%} of artifacts experience reproduction failures at least once.
           & Researchers should report environment information, such as the Qiskit version and the date on which the experiments were conducted. When artifacts are modified to make them executable, researchers should document these modifications.
           & \cmark                                                                                                                                                                                                                                                          \\
    \cmidrule(l){2-4}

           & \textbf{PF2:} \underline{15.3\%} of artifacts break multiple times. \newline
    \textbf{OF2:} \underline{83.8\%} of artifacts (i.e., all broken artifacts) break multiple times.
           & Dataset maintainers should regularly check reproducibility and promptly migrate affected artifacts to new APIs when maintaining quantum software defect datasets built on actively evolving frameworks such as Qiskit.
           & \xmark                                                                                                                                                                                                                                                          \\
    \midrule

    \multirow{2}{=}{\raggedright RQ2: Root causes of reproduction failures}
           & \textbf{PF3:} \underline{93.9\%} of root causes are dependency-related (e.g., Maven TLS Failure). \newline
    \textbf{OF3:} \underline{93.6\%} of root causes are dependency-related (Source-related 82.0\% + Library-related 11.6\%).
           & Maintenance of quantum software defect datasets should prioritize dependency management (e.g., version pinning, compatibility tracking, and dependency caching), as in classical ones.
           & \cmark                                                                                                                                                                                                                                                          \\
    \cmidrule(l){2-4}

           & \textbf{PF4:} \underline{93.9\%} of broken artifacts can be handled by dependency updates alone, without modifying the source. \newline
    \textbf{OF4:} \underline{4.6\%} of broken artifacts can be handled by dependency updates alone; the rest require source modifications or are not patchable.
           & Automated code-migration techniques that follow Qiskit API evolution are a promising research direction for maintaining quantum software defect datasets.
           & \xmark                                                                                                                                                                                                                                                          \\

    \bottomrule
  \end{tabular}
\end{table*}

Based on the previous two RQs, this section discusses to what extent the four findings of Zhu et al.~\cite{zhu2023icse} on classical software defect datasets generalize to the quantum software defect dataset Bugs4Q.
\cref{tab:replication-discussion} shows their findings (PF1--PF4) with ours (OF1--OF4) for each RQ, together with the implications drawn from each comparison.
Two of the four findings (PF1/OF1 and PF3/OF3) are consistent between the classical and quantum settings, whereas the remaining two (PF2/OF2 and PF4/OF4) diverge.
We discuss the factors behind the similarities or differences and articulate implications (see \faHandPointRight[regular] icon) for the broader research community.

\textbf{PF1 vs. OF1 (\cmark): Both classical and quantum software datasets suffer from reproduction failures over time.}
Zhu et al. reported that 62.6\% of BugSwarm~\cite{tomassi2019icse} artifacts experienced at least one reproduction failure over 13 months (PF1).
Similarly, we found that 83.8\% of Bugs4Q artifacts did so over three years (OF1).
This agreement suggests that reproduction failures are a common threat to defect datasets, regardless of whether they target classical or quantum software.
A possible explanation is that, in both types of defect datasets, artifact reproducibility depends on external components whose evolution is outside the control of dataset maintainers.

\implication{For researchers}, it is important to recognize that reproducibility depends on both time and the execution environment.
Thus, studies using Bugs4Q should report the dataset snapshot, execution environment (e.g., the Qiskit version), and which artifacts could not be reproduced.
When artifacts are modified to make them executable, the applied modifications should be documented in the paper or the replication package.

\textbf{PF2 vs. OF2 (\xmark): Reproduction failures persist more strongly in quantum software datasets.}
Zhu et al. reported that only 15.3\% of BugSwarm artifacts experienced reproduction failures multiple times (PF2).
In contrast, in Bugs4Q, all 31 artifacts that experienced reproduction failures (83.8\% of the dataset) did so multiple times across snapshots (OF2).
This divergence may be explained by the scale of breaking changes introduced by Qiskit major version upgrades.
In Bugs4Q, artifacts are exposed to major Qiskit version transitions (i.e., 0.x$\rightarrow$1.x and 1.x$\rightarrow$2.x), which involved broad API restructuring as documented in the official migration guides~\cite{qiskit2024migration,qiskit2025migration}.
Therefore, once an artifact becomes unreproducible, it may remain unreproducible across later snapshots unless the code is explicitly migrated to the new APIs.

\implication{For dataset maintainers}, one-shot patches are insufficient for maintaining quantum software datasets built on actively evolving frameworks such as Qiskit.
They need continuous maintenance, including promptly migrating affected artifacts to new APIs when reproduction failures are detected.

\textbf{PF3 vs. OF3 (\cmark): Dependency-related reproduction failures dominate in both settings.}
Zhu et al. reported that 93.9\% of reproduction failures in classical defect datasets are dependency-related, such as Maven TLS failures and Unavailable Gradle Plugin (PF3).
We observed a similar proportion in Bugs4Q: 93.6\% of reproduction failures are dependency-related, comprising Source-related failures (82.0\%) and Library-related failures (11.6\%) (OF3).
This agreement suggests that dependency management is essential for defect datasets in general, because artifacts inherently depend on specific versions of external libraries to expose the originally reported bugs.

\implication{For dataset maintainers}, maintenance of quantum software defect datasets should incorporate dependency-aware practices such as version pinning, compatibility tracking across the surrounding package ecosystem, and dependency caching~\cite{zhu2023icse}, as these practices are also important for classical datasets.

\textbf{PF4 vs. OF4 (\xmark): Dependency patches alone are not sufficient to restore reproducibility for quantum software defect datasets.}
While the dependency-aware practices previously discussed are necessary, they are not sufficient for our setting.
Zhu et al. reported that 93.9\% of reproduction failures can be recovered by dependency updates alone, without modifying the source code (PF4).
In Bugs4Q, however, only 4.6\% of reproduction failures are addressable by dependency updates alone; the remaining 95.4\% require source-level modifications, such as migrating import paths and API invocations (OF4).
This divergence reflects the rapid evolution of the Qiskit ecosystem, which has repeatedly removed or restructured public APIs at major version boundaries, making source-code references themselves stale.

\implication{For researchers}, developing automated code-migration techniques that follow Qiskit API evolution is an important research direction, since manual source-level patching does not scale.
Our manually patched dataset, Bugs4Q-Robust, provides a basis for this direction by offering concrete migration patches.
Future work could further combine these patches with knowledge mined from Qiskit migration guides to automate the maintenance of quantum defect datasets.

\section{Threats to Validity}
\label{sec:threat}

\textbf{Construct Validity.}
Among the three reproducibility criteria, Cause Match requires manual inspection because determining whether different runtime logs stem from the same underlying cause cannot be automated.
In our study, this inspection was performed solely by the first author, which may introduce subjectivity.
We documented the inspection criteria and made all classification results publicly available in our replication package for external verification.
Another threat concerns the fidelity of environment reproduction.
Although we align the versions of Qiskit-related packages with each snapshot, other dependencies may not exactly match the dependency state at the corresponding time.
Thus, our Docker images may not perfectly reproduce the exact environment of each snapshot.
We mitigate this threat by pinning the versions of most dependencies in the requirements files.
Since the potentially mismatched dependencies are not Qiskit-related packages, we expect their impact on our results to be limited.

\textbf{Internal Validity.}
Two design choices may affect the results: (1) the number of executions per version and (2) the package-version selection policy in Pinned-stack.
For the former, we execute each version 30 times following empirical guidance for randomized algorithms~\cite{arcuri2011icse}.
For the latter, we select package versions based on official Qiskit release notes to reflect the intended package combinations at each snapshot.
Different choices could yield different values of reproducibility, but the overall trends should remain consistent.
Manual classification and patching in RQ2 may also affect our results.
The classification is performed independently by two authors, achieving substantial agreement, with disagreements resolved through discussion to reach a final consensus.
Patches are kept minimal with the patched code and logs inspected for unrelated changes.
The replication package includes the patched artifacts and runtime logs for external verification.

\textbf{External Validity.}
Our study targets a single dataset (Bugs4Q) and a single framework (Qiskit),
which may limit the generalizability of our findings to other quantum frameworks or defect datasets.
This is inherent to the current state of the field: Qiskit is a widely used framework, and Bugs4Q is, to our knowledge,
the only publicly available quantum defect dataset that bundles real-world buggy programs with their fixed versions and bug-reproducing tests.
Other frameworks such as Cirq, PennyLane, and Q\# may exhibit different API-evolution patterns. Validating our findings on additional quantum software defect datasets built on these frameworks remains future work.
Our analysis covers 21 core-library versions across the 0.x--2.x series, ending at v2.3.1,
so the observed trends may not hold for future Qiskit versions. Evaluating and maintaining \bqp under such versions is a direction for follow-up work.

\section{Related Work}
\label{sec:related_work}

This section reviews related work in two areas: software defect datasets and bugs in quantum programs.

\subsection{Software Defect Datasets and Their Reproducibility}
Software defect datasets are essential for evaluating software engineering techniques.
A survey~\cite{zhu2026csur} reviewed 151 software defect datasets using January 2025 as the cutoff date.
Zhu et al.~\cite{zhu2026csur} reported that over 60\% of the datasets target general-purpose software, including widely used benchmarks such as Defects4J~\cite{just2014issta}, ManyBugs~\cite{le2015tse}, and BugSwarm~\cite{tomassi2019icse}.
The survey reported three defect datasets related to quantum computing~\cite{paltenghi2022oopsla,zhao2023qsw,zhao2023jss}.
Among them, Bugs4Q~\cite{zhao2023jss} is the only dataset that targets bugs in user-written quantum programs and provides fixed programs and test code for reproducing the bugs.
In contrast, the other two datasets~\cite{paltenghi2022oopsla,zhao2023qsw} focus on bugs in quantum computing platforms or frameworks.
While our literature search further found additional quantum software defect datasets~\cite{ishimoto2025ease,usandizaga2026arxiv}, we exclude these datasets from our study because their bugs are synthetically injected through quantum circuit mutations~\cite{fortunato2022tqe}, rather than real-world bugs.

Prior studies~\cite{zhu2023icse,aguilar2023scam} have examined whether software defect datasets remain reproducible after their construction.
Zhu et al.~\cite{zhu2023icse} investigated the reproducibility of five Java defect datasets, such as Defects4J~\cite{just2014issta}.
They showed that defect datasets are not immune to reproduction failures over time, even when the source code of the artifacts remains unchanged.
Aguilar et al.~\cite{aguilar2023scam} revisited BugsInPy~\cite{widyasari2020fse}, a Python defect dataset.
They argued that \texttt{requirements.txt} is not enough for reproducing bugs and that more detailed environment specifications, such as container images, should be provided.
This observation is consistent with our finding that execution environments and dependency configurations strongly affect reproducibility of Bugs4Q.

\subsection{Bugs in Quantum Programs}
Prior studies~\cite{zhao2021qse,paltenghi2022oopsla,luo2022saner,zhao2023qsw} have shown that quantum-specific concepts, such as gate operations and measurement, often lead to bugs in quantum programs.
For instance, Luo et al.~\cite{luo2022saner} conducted an empirical study of 96 real-world quantum program bugs and their fixes.
Their results show that more than 80\% of the bugs are quantum-specific, meaning that fixing them requires knowledge of quantum computing.
In contrast, Zappin et al.~\cite{zappin2025icse} reported a trend when studying recurring issues faced by developers of hybrid quantum-classical (HQC) applications, a more realistic setting in the noisy intermediate-scale quantum (NISQ) era where quantum and classical components are integrated to overcome current hardware limitations.
They found that only 16\% of 483 issues are quantum-specific, whereas the majority arise on the classical side or at the boundary between the two.

Our finding aligns with that of Zappin et al.:
quantum-specific factors are not the dominant cause of reproduction failures in quantum defect datasets, accounting for only 5.1\% of reproduction failures of Bugs4Q.
This alignment is plausible because both studies target realistic settings, where artifacts depend on broader software ecosystems rather than on quantum-specific logic alone.
Our results further suggest that, in the context of defect dataset reproducibility, the key factor is whether it depends on a software ecosystem that frequently introduces breaking changes (e.g., the Qiskit ecosystem) rather than whether a dataset targets classical or quantum software.

\section{Conclusions}
\label{sec:conclusion}
We investigated the reproducibility of Bugs4Q~\cite{zhao2023jss}, a representative quantum software defect dataset.
Our experiments covered 77{,}700 executions of 37 artifacts across 21 Qiskit core-library versions spanning three years.
The results showed that the reproducibility of Bugs4Q dropped sharply after major Qiskit upgrades (e.g., from 62.2\% on Qiskit v0.20.1 to 16.2\% on v2.3.1, the latest version as of April 1, 2026).
Furthermore, 83.8\% of the artifacts experienced reproduction failures at least once.
Through manual investigation of 543 reproduction failures, we found that 93.6\% are dependency-related, consistent with findings on classical software defect datasets~\cite{zhu2023icse}.
However, unlike classical datasets, only 4.6\% of reproduction failures of Bugs4Q can be addressed by dependency updates alone.
Guided by these findings, we curated \bqp, a patched version of Bugs4Q for Qiskit v2.3.1, which raises the reproducibility by 51.4 percentage points on average.
Our investigation suggests that maintaining quantum defect datasets can require more than preserving executable environments; source code and test oracles may also need to evolve with the underlying quantum software framework.

\section{Data Availability} \label{sec:data_availability}
All data and code used in this study are available at the following repository: \ReplicationURL

\section*{Acknowledgments}
This research was partially supported by JSPS KAKENHI Japan (Grant Number: JP26H02500, JP25K15056, JP25K03102, and JP24H00692).

\bibliographystyle{IEEEtran}
\bibliography{reference}

\end{document}